\documentstyle[preprint,eqsecnum,aps,prb]{revtex}

\def\et{{\it et al.}}
\def\gb{{\cal G_B}}
\def\gs{{\cal G_S}}
\def\prb{Phys. Rev. B}

\def\fns{\footnotesize}

\begin{document}
\draft
\preprint{Proccedings of CLACSA. Cancun, Mexico, Sep. 21--28, 1994}

\title{Two--dimensional Bulk Bands and Surface Resonances
Originated from (100) Surfaces
of III--V Semiconductor Compounds}

\author{D. Olgu\'{\i}n,
R. de Coss$^\dagger$,
and R. Baquero}

\address{Departamento de F\'{\i}sica\\
Centro de Investigaci\'on y de Estudios Avanzados del IPN\\
Apartado Postal 14--740, 07000, M\'exico, D.F.}

\address{$^\dagger$Departamento de F\'\i sica Aplicada\\
Centro de Investigaci\'on y de Estudios Avanzados del IPN--Unidad M\'erida,\\
Apartado Postal 73, Cordemex 97310, M\'erida, Yucat\'an, M\'exico.
}
\maketitle

\begin{abstract}
We have calculated the electronic band structure 
of the (100) surface of the III--V zinc blende semiconductor 
compounds, 
using the 
standard
tight binding method and the surface 
Green's function matching method. 
We have found
that the creation of the surface gives place to new states 
in the electronic structure: surface resonances and two dimensional 
bulk states. 
The two dimensional bulk states are of the same character of those
reported recently
in CdTe(100) [Phys. Rev. {\bf 50}, 1980 (1994)]. We analyze
the states in the valence band
region and compare with photoemission spectroscopy data. 
\end{abstract}

\section{INTRODUCTION}

It is very well known that the creation of a surface gives origin to new 
electronic states in the material, these states are called {\it surface 
states}. The surface states are in the gaps of the projected bulk bands 
and have two dimensional (2D) character. Recently we have given a 
theoretical description 
of the $B-4$ band in CdTe(100) (1), and we found that this band 
is of 2D character but does exist in atomic layers away from the surface,
it is in this sense a {\it bulk band}, not a surface state. The appearance
of this band is a consequence of the creation of the (100) surface.
In the present work we made a systematic study of the valence band 
electronic structure of the (100) surface in the III--V zinc blende 
semiconductors, with the purpose to determine if the 
character of the 2D bulk band found in CdTe(100) is universal.

An experimental technique capable of identifying absolute critical point 
energies and the dispersion of valence band is angle resolved 
photoelectron spectroscopy (ARPES). ARPES has been very successful in the 
investigation of IV, III--V, II--VI, and IV--VI semiconductor 
compounds (2).
The III--V semiconductor compounds have been subject of extensive studies 
in the past, in particular the (100) surface of the zinc blende 
compounds GaAs (3),
GaSb (4), and 
InSb (5). 
The experimental reports give account of the surface states in the 
valence band and in some cases other structures no reported previously. 
It is in the last point that our work is focussed. 
Using the tight 
binding method (6) (TBM) 
with the surface Green's function 
matching method (7) (SGFM) we calculate the bulk band structure 
projected on the (100) surface and we compare it with the 
experimental data. The 
SGFM method allows us to determine the origin of all those 
states (1,7). \\

\section{THE METHOD}

In order to study the (100) surface we use the SGFM method and calculate the 
(100) bulk projected (${\cal G_B}$) and the (100) surface projected 
(${\cal G_S}$) Green's function.
From $\gb$\ we can calculate the effects on the band structure coming from 
the hard wall effect (in close analogy to the hard core term in 
scattering theory, see Ref. (8)) and from $\gs$\ we obtain the surface band 
structure. To calculate these Green's functions we use the known formula 
\begin{equation}
\gs^{-1}=(E-H_0)-H_{01}T,
\end{equation}
\begin{equation}
\gb^{-1}=\gs^{-1}-H^\dagger_{01}\widetilde T.
\end{equation}
The way in which the Hamiltonian matrices $(H_{ij})$ are related to the TBM is 
described in detail in Ref. (9). $T$ and $\widetilde T$ are the transfer 
matrices depending only on $H_{00}$, $H_{01}$, and $H^\dagger_{01}$, the 
principal layer projected Hamiltonian matrices (for further discussion 
see Refs. (7, 8)). \\

\section{RESULTS AND DISCUSSION}

By looking at the poles of the real part of the Green's function, 
we found the eigenvalues for each value of {\bf k}, a vector 
in the Brillouin zone. In this way we have calculated the surface states
and the 2D bulk states in the (100) surface of the III--V semiconductor
compounds.
In general, we obtain two surface resonances, the first one with 
anion terminated 
surface character and the second one with cation terminated surface 
character. We have three 2D bulk states 
and one bulk resonance state. 
These states appear in the band structure due to the breaking of 
the crystal translational symmetry in the (100) direction.
The creation
of the surface gives origin to new states in the electronic structure,
both surface
and bulk states, as those discussed in a recent work for
II--VI compounds (1, 10--12).
We present a brief discussion of our results in the GaAs(100), GaSb(100), 
and InSb(100) compounds. Here we only discuss 
the GaAs(100) surface and for the other compounds we make a brief 
mention of the principal characteristics found in our calculation,
a more extensive discussion will be published elsewhere.
Figure 1 shows our calculated band structure, in the $\Gamma-X$ direction, 
for GaAs(100). 
In the figure we show the bulk band structure (solid lines), the surface 
resonances (triangles), and the 2D bulk states (points).
\begin{figure}[!h]
\input epsf
\vspace*{-.1in}
\begin{minipage}[h]{4 true in}
\vspace*{-0.5 true cm}
\hfill
\epsfbox{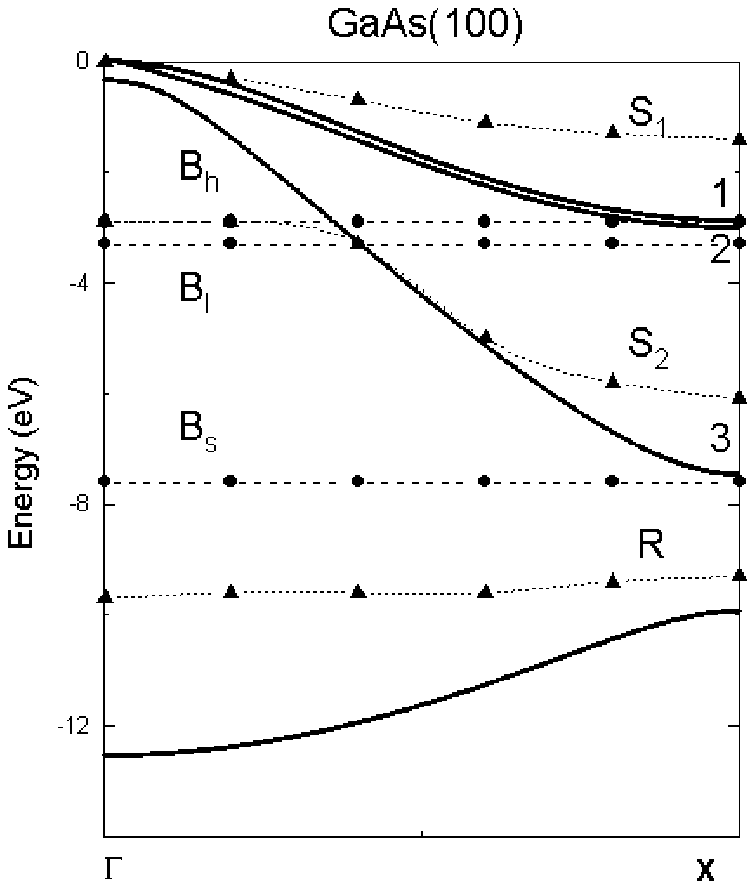}
\vspace*{-.25 true cm}
\end{minipage}
\centerline{\fns{\bf FIGURE 1.} 
Projected (100) surface band structure for GaAs. We show the bulk bands,}
\parskip-6pt
{\fns labeled 1, 2, and 3 (solid lines), the surface resonances 
(triangles), 
and the 2D bulk states (points). The broken lines are only guide to the eye.}
\end{figure}
\vspace{.3cm}
We identify the bulk bands calculated by Chelikowsky and 
Cohen (13), by the solid lines. We found these bulk bands after the 
diagonalization of the bulk tight binding Hamiltonian with the parameters 
of the Ref. (14).
The surface resonances and the 2D bulk  
states have been obtained looking at the poles of the real part of the 
surface Green's function (RPSGF) and the poles of the real part of 
the bulk Green's function (RPBGF), respectively, in the same way 
as is discussed in Refs. (1,10).
The $S1$ surface resonance, labeled $S$ by Olde \et\ (3)
(see Fig. 1 from Ref. (3)), appears due 
to the creation of the surface in the material and it is characteristic 
of the anion terminated surface. The state begins at the top 
of the valence band in $\Gamma$ and shows some dispersion in 
the $\Gamma-X$ direction. The $S2$ surface resonance, 
that shows larger dispersion 
than the $S1$ resonance, begins at --2.9 eV in $\Gamma$
and reaches $X$ at --6.1 eV, approximately. This state is 
characteristic of the cation terminated surface.
On the other hand, we obtain three 2D bulk states, $B_h,\ B_l$ and $B_s$, 
in the 
GaAs(100) surface. The $B_h$ state is located at --2.9 eV, 
the $B_l$ state at 
--3.3 eV, and the $B_s$ state
at --7.6 eV, approximately. These states do not 
show dispersion. The 2D bulk states appear for both, the anion terminated 
and cation terminated, surfaces. 
Finally, we have the $R$ state at --9.9 eV, this state shows slight 
dispersion and we identify 
it as a bulk resonance state. The state is a bulk resonance because 
this state mixes with the projected band structure, the projection 
of the bulk band
$E_{n,{\bf k}}$ for each {\bf k} in the surface Brillouin zone, at $\Gamma$ 
and goes to the forbidden gap for {\bf k} values near $X$ (15).


According to this, we suggest the next theoretical description of 
the band structure reported by Olde \et\ (3): 
the $S1$ surface resonance could be the $S$ experimental band, one of our 
2D bulk states,
$B_h$ or $B_l$, could be the $S'$ band, the $S2$ surface
resonance 
could be identified as the $c$ band,
and the $B_s$ state with the $d$ band. From the experimental report
it is not possible to see if the bulk 
resonance state $R$ has an experimental counterpart. 


The calculated band structure for GaSb(100) and InSb(100) shows the same
pattern as GaAs(100). We obtain the surface resonances $(S1$ and $S2)$, 
the 2D bulk states $(B_h,\ B_l$ and $B_s)$, as well as the bulk 
resonance state $(R)$. We could made a similar identification, as in the 
GaAs(100) case, of our calculated states using the experimental reports 
of Franklin {\it et al.} (4) and Middelmann {\it et al.} (5) for GaSb(100) 
and InSb(100), respectively. However, we do not present further discussions
here. A more extensive discussion will be published elsewhere.

\section{CONCLUSIONS}

We have studied the electronic structure of the 
(100) surface of the zinc blende semiconductor compounds GaAs, 
GaSb, and InSb that have been experimentally obtained using 
angle--resolved photoelectron spectroscopy. We used 
an empirical 
tight binding description 
and 
the surface Green's 
function matching method 
to obtain 
the bulk bands and the electronic structure projected on the surface. 
From the Green's function we have established the character of the 
surface-- and bulk--states.
An important point that we obtained from our calculation is the 
appearance of three 2D bulk states and the known surface resonances 
of these surfaces. The 2D
bulk states appear in the electronic structure due to the breaking 
of the crystal translational symmetry in the (100) direction, 
as we have shown for II--VI compounds.\\

\section*{ACKNOWLEDGMENTS}
The authors would like to thank Dr. M. L\'opez for stimulating and critical 
comments.
One of 
us (D.O.) acknowledges the support of CONACYT, Mexico.\\

\vspace{0.3cm}
\centerline{\large\bf REFERENCES}
\vspace{0.3cm}

\begin{enumerate}
\item Olgu\'\i n D., and Baquero R., \prb\ {\bf 50}, 1980--1983 
(1994).
\item Leckley R. C. G., and Riley J. O., {\it Critical Review in 
Solid State and Matt. Scie.} {\bf 17}, 307--352 (1992).
\item Olde J., {\it et al.}, \prb\ {\bf 41}, 9958--9965 (1990).
\item Franklin G. E., {\it et al}, \prb\ {\bf 41}, 
12 619--12 627 
(1990).
\item Middelmann H. U., , Sorba L., , Hinkel V., and Horn K., \prb\ 
{\bf 34}, 957--962 (1986).
\item Slater J. C., and Koster G. F., {\it Phys. Rev.} 
{\bf 94}, 1498--1524 
(1954).
\item Garc\'\i a--Moliner F., and Velasco V., {\it Prog. Surf. Scie.} 
{\bf 21}, 93--162 (1986). 
\item Garc\'\i a--Moliner F., and Velasco V., {\it Theory 
of Single and Multiple Interfaces} (World Scientific, Singapore 1992).
\item Baquero R., and Noguera A., {\it Rev. Mex. F\'\i s.}
{\bf 35}, 638--647 (1989).
\item Olgu\'\i n D., and Baquero R., \prb\ {\bf 51}, 16 981--16 987
(1995).
\item Niles D., and H\"ochst H., \prb\ {\bf 43}, 1492--1499 
(1991).
\item Gawlik K. -U., \et , {\it Acta Physica Pol. A}
{\bf 82}, 355--361 (1992).
\item Chelikowsky J. R., and Cohen M. L., \prb\ {\bf 
14}, 556--582 (1976).
\item Priester C., Allan G., and Lanno M., \prb\ {\bf 
37}, 8519--8522 (1988).
\end{enumerate}

\end{document}